\documentclass[twocolumn,prl,preprintnumbers,amsmath,amssymb,superscriptaddress]{revtex4}
\usepackage{graphicx}
\usepackage{epsfig}
\usepackage{amsmath,amssymb,bm}
\usepackage{color}
\usepackage[switch]{lineno}
\usepackage[normalem]{ulem}

\newcommand{\tHe}{$^{3}$He}
\newcommand{\beginsupplement}{%
        \setcounter{table}{0}
        \renewcommand{\thetable}{S\arabic{table}}%
        \setcounter{figure}{0}
        \renewcommand{\thefigure}{S\arabic{figure}}%
     }

\begin{document}

\begin{titlepage}
This manuscript has been authored by UT-Battelle, LLC under Contract No. DE-AC05-00OR22725 with the U.S. Department of Energy. The United States Government retains and the publisher, by accepting the article for publication, acknowledges that the United States Government retains a non-exclusive, paid-up, irrevocable, world-wide license to publish or reproduce the published form of this manuscript, or allow others to do so, for United States Government purposes. The Department of Energy will provide public access to these results of federally sponsored research in accordance with the DOE Public Access Plan(http://energy.gov/downloads/doe-public-access-plan).
\newpage
\end{titlepage}

\title{Direct observation of the Higgs amplitude mode in a two-dimensional quantum antiferromagnet near the quantum critical point}

\author{Tao Hong}
\email[Electronic address: ]{hongt@ornl.gov}
\affiliation{Quantum Condensed Matter Division, Oak Ridge National Laboratory, Oak Ridge, Tennessee 37831, USA}
\author{Masashige Matsumoto}
\affiliation{Department of Physics, Shizuoka University, Shizuoka 422-8529, Japan}
\author{Yiming Qiu}
\affiliation{National Institute of Standards and Technology, Gaithersburg, Maryland 20899, USA}
\author{Wangchun Chen}
\affiliation{Department of Materials Science and Engineering, University of Maryland, College Park, Maryland 20742, USA}
\author{Thomas R. Gentile}
\affiliation{National Institute of Standards and Technology, Gaithersburg, Maryland 20899, USA}
\author{Shannon Watson}
\affiliation{National Institute of Standards and Technology, Gaithersburg, Maryland 20899, USA}
\author{Firas F. Awwadi}
\affiliation{Department of Chemistry, The University of Jordan, Amman 11942, Jordan}
\author{Mark M. Turnbull}
\affiliation{Carlson School of Chemistry and Biochemistry, Clark University, Worcester, Massachusetts 01610, USA}
\author{Sachith E. Dissanayake}
\affiliation{Quantum Condensed Matter Division, Oak Ridge National Laboratory, Oak Ridge, Tennessee 37831, USA}
\author{Harish Agrawal}
\affiliation{Instrument and Source Division, Oak Ridge National Laboratory, Oak Ridge, Tennessee 37831, USA}
\author{Rasmus Toft-Petersen}
\affiliation{Helmholtz-Zentrum Berlin f$\ddot{\rm u}$r Materialien und Energie, D-14109 Berlin, Germany}
\author{Bastian Klemke}
\affiliation{Helmholtz-Zentrum Berlin f$\ddot{\rm u}$r Materialien und Energie, D-14109 Berlin, Germany}
\author{Kris Coester}
\affiliation{Lehrstuhl f$\ddot{u}$r Theoretische Physik I, TU Dortmund, D-44221 Dortmund, Germany}
\author{Kai P. Schmidt}
\affiliation{Lehrstuhl f$\ddot{u}$r Theoretische Physik I, Staudtstrasse 7, Universit$\ddot{a}$t Erlangen-N$\ddot{u}$rnberg, D-91058 Germany}
\author{David A. Tennant}
\affiliation{Quantum Condensed Matter Division, Oak Ridge National Laboratory, Oak Ridge, Tennessee 37831, USA}

\begin{abstract}
Spontaneous symmetry-breaking quantum phase transitions play an essential role in condensed matter physics\cite{Sondhi97:69,Sachdev99,Vojta03:66}. The collective excitations in the broken-symmetry phase near the quantum critical point can be characterized by fluctuations of phase and amplitude of the order parameter. The phase oscillations correspond to the massless Nambu$-$Goldstone modes whereas the massive amplitude mode, analogous to the Higgs boson in particle physics\cite{Higgs64:13,Guralnik64:13}, is prone to decay into a pair of low-energy Nambu$-$Goldstone modes in low dimensions\cite{Chubukov94:49,Sachdev99,Sachdev99:59}. Especially, observation of a Higgs amplitude mode in two dimensions is an outstanding experimental challenge. Here, using the inelastic neutron scattering and applying the bond-operator theory, we directly and unambiguously identify the Higgs amplitude mode in a two-dimensional \emph{S}=1/2 quantum antiferromagnet C$_9$H$_{18}$N$_2$CuBr$_4$ near a quantum critical point in two dimensions. Owing to an anisotropic energy gap, it kinematically prevents such decay and the Higgs amplitude mode acquires an infinite lifetime.
\end{abstract}

\maketitle

The Higgs boson appears as the amplitude fluctuation of the condensed Higgs field in the Standard Model of particle physics. Since its discovery, there has been much interest in searching for similar Higgs boson-like particles in condensed matter physics, such as in superconductors\cite{Sooryakumar-1980,Litlewood-1982,Matsunaga-2014,Sherman-2015}, charge-density-wave systems\cite{Tsang-1976,Pouget-1991}, ultracold bosonic systems\cite{Endres12:487}, and antiferromagnets\cite{Ruegg08:100,Merchant14:10,Grenier15:114}. Strictly speaking, only superconductors are analogous to the particle physics for the point that the gauge field (photon) coupling to the condensate acquires its mass (Meissner effect) by means of the symmetry breaking (Anderson-Higgs mechanism). In a broad sense, nevertheless, the excitation mode of the amplitude fluctuation of the order parameter is also termed as "Higgs amplitude mode" in condensed matter physics\cite{Pekker15:6}. These works provided new insights about the fundamental theories underlying the exotic materials.

The Higgs amplitude mode is expected in the proximity of a quantum critical point (QCP) but can decay into a pair of low-energy Nambu-Goldstone modes, which makes it experimentally difficult to detect. In three-dimensional (3D) systems, where the QCP is a Gaussian fixed point, the Higgs amplitude mode is well defined near QCP\cite{Lohofer17:118}. In contrast, in the two-dimensional (2D) case, where the longitudinal susceptibility becomes infrared divergent near QCP, it has been debated whether the Higgs amplitude mode may not survive or it is still visible in terms of a scalar susceptibility\cite{Chubukov94:49,Sachdev99:59,Podolsky-2011,Pollet12:109,Gazit-2013,Chen-2013,Lohofer15:92}. Indeed, the Higgs amplitude mode in 2D was evidenced by the scalar response for an ultracold atomic gas near the superfluid to Mott-insulator transition\cite{Endres12:487} and for a disordered superconductor close to the superconductor-insulator transition\cite{Sherman-2015} although the observed spectral function is heavily damped. Note that when the Nambu$-$Goldstone modes become gapped, there is no such physical infrared singularity. In the following paper, we will demonstrate observation of the sharp Higgs amplitude mode through the longitudinal response being such case in an \emph{S}=1/2 2D coupled-ladder compound C$_9$H$_{18}$N$_2$CuBr$_4$ (abbreviated as DLCB) in the vicinity of QCP in two dimensions.

\begin{figure}
\includegraphics[width=8cm,bbllx=85,bblly=225,bburx=520,bbury=675,angle=90,clip=]{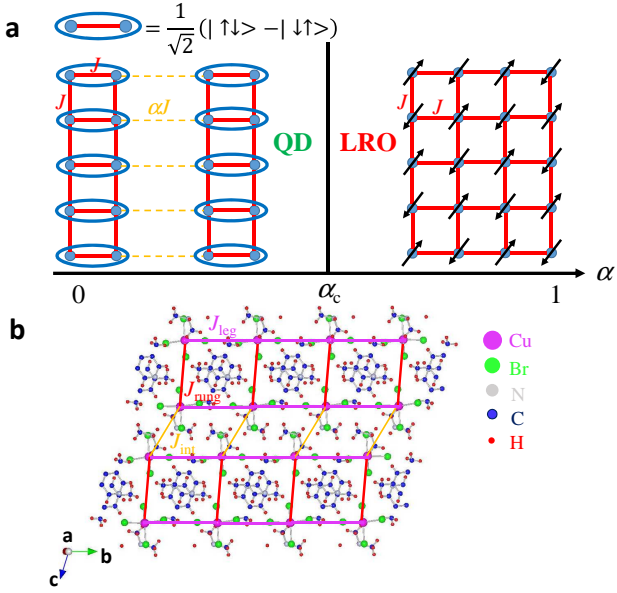}
\caption{\textbf{The two-dimensional coupled two-leg spin ladder antiferromagnet.} \textbf{a}, Schematic diagram of coupled two-leg square spin ladders, where the ground state can be tuned by the inter-ladder coupling $\alpha J$ from the quantum disordered (QD) phase, through a quantum critical point $\alpha_c$ in two dimensions, to the long-range magnetically ordered (LRO) phase. Blue circles stands for the spin-1/2 magnetic ions. The ellipses represent a singlet valence bond of spins. \textbf{b}, The molecular two-leg ladder structure with the leg direction along the crystalline b axis and a two-dimensional model for the magnetic interactions in C$_9$H$_{18}$N$_2$CuBr$_4$. Pink, red, and yellow bonds indicate the nearest neighbor leg, rung, and inter-ladder exchange constants, respectively.}
\label{fig1}
\end{figure}

The quantum \emph{S}=1/2 Heisenberg antiferromagnetic two-leg spin ladder is one of the cornerstone models in low-dimensional quantum magnetism\cite{Barnes93:47,Dagotto96:271}. In the one-dimensional limit of isolated spin-1/2 ladders, the ground state consists of dressed valence-bond singlets on each rung of the ladder. Interestingly, the ground state as shown in Fig.1a can be tuned by the inter-ladder coupling from the quantum disordered (QD) state, through the QCP, to the renormalized classical regime of a long-range magnetically ordered (LRO) state\cite{Troyer97:55,Normand97:56}. In the QD phase, the magnetic excitations are triply degenerate magnons with a spin gap energy $\Delta$ which vanishes on approach to the QCP. In the LRO phase, the triplet modes evolve into two gapless Nambu-Goldstone modes reflecting spin fluctuations perpendicular to the ordered moment, accompanied by a longitudinal mode reflecting spin fluctuations along the ordered moment. The latter mode has a gap which grows continuously with the moment and is analogous to the Higgs amplitude mode. Such a longitudinal mode is usually unstable and decays into a pair of the transverse modes as observed in the \emph{S}=1/2 coupled Heisenberg chain compound KCuF$_3$ and has a finite lifetime\cite{Lake00:85,Lake05:71}.

In our previous work\cite{Hong14:89,Hong17:08}, we have shown that a metal-organic compound DLCB is a unique spin ladder material where the inter-ladder coupling is sufficiently strong to drive the system into the magnetically ordered phase. Figure~1b shows the molecular two-leg ladder structure of DLCB. The collinear magnetic structure was determined by the unpolarized neutron diffraction technique and the staggered moments point along the c* axis ($\equiv$$\hat{z}$) with a reduced moment size about 0.4 $\mu_{\rm B}$\cite{Hong14:89}. A minimal 2D spin interacting model was proposed based on the crystal structure and the corresponding spin Hamiltonian can be written as:
\begin{eqnarray} \label{ladeqn}
H &=&  J_{\rm rung}\sum_{l,i}(\lambda{\bf S}^{x}_{l_1,i}{\bf S}^{x}_{l_2,i}+\lambda{\bf S}^{y}_{l_1,i}{\bf S}^{y}_{l_2,i}+{\bf S}^z_{l_1,i}{\bf S}^z_{l_2,i}) \\ \nonumber
&+&J_{\rm leg}\sum_{l,i}(\lambda{\bf S}^{x}_{l_1,i}{\bf S}^{x}_{l_1,i+1}+\lambda{\bf S}^{y}_{l_1,i}{\bf S}^{y}_{l_1,i+1}+{\bf S}^z_{l_1,i}{\bf S}^z_{l_1,i+1} \\ \nonumber
&+& \lambda{\bf S}^{x}_{l_2,i}{\bf S}^{x}_{l_2,i+1}+\lambda{\bf S}^{y}_{l_2,i}{\bf S}^{y}_{l_2,i+1}+{\bf S}^z_{l_2,i}{\bf S}^z_{l_2,i+1}) \\ \nonumber &+&J_{\rm int}\sum_{l,i}(\lambda{\bf S}^{x}_{l_1,i}{\bf S}^{x}_{(l-1)_2,i}+\lambda{\bf S}^{y}_{l_1,i}{\bf S}^{y}_{(l-1)_2,i}+{\bf S}^z_{l_1,i}{\bf S}^z_{(l-1)_2,i} \\ \nonumber
&+&\lambda{\bf S}^{x}_{l_2,i}{\bf S}^{x}_{(l+1)_1,i}+\lambda{\bf S}^{y}_{l_2,i}{\bf S}^{y}_{(l+1)_1,i}+{\bf S}^{z}_{l_2,i}{\bf S}^{z}_{(l+1)_1,i}),
\end{eqnarray}
where \emph{l} indexes the site of a ladder, \emph{i} for rungs, and 1, 2 for each leg. $J_{\rm rung}$, $J_{\rm leg}$, and $J_{\rm int}$ are the nearest-neighbor rung, leg, and inter-ladder exchange constants. $\lambda$ (between 0 and 1) specifies an exchange anisotropy with $\lambda$=0 or 1 being the limiting case of Ising or Heisenberg interactions. Indeed, the observed magnon dispersions can be described by this model quantitatively\cite{Hong14:89}. An observed spin gap energy $\Delta$=0.32(3) meV is owing to a small Ising anisotropy. Further evidence for the two-dimensionality is provided by the measurement of the inter-layer dispersion as described in the Supplementary Information.

Importantly, the analysis of the spin Hamiltonian indicates that DLCB is close to a QCP\cite{Hong14:89}, making DLCB an ideal \emph{S}=1/2 quantum antiferromagnet to investigate the Higgs amplitude mode. The more detailed examination of this QCP can be found in the Supplementary Information. In the easy-axis case, the U(1) symmetry is not spontaneously broken in the ordered phase and the two branches of transverse modes (TMs) would have equal gap energies as predicted by the spin-wave theory. Therefore, if the longitudinal mode (LM)/the Higgs amplitude mode is sufficiently close to TMs, it could become kinematically unable to decay in this region and thus acquire an infinite lifetime.

Since TMs and LM (Higgs amplitude mode) could be degenerate within the instrumental resolution in DLCB, we carried out a follow-up unpolarized inelastic neutron scattering (INS) study in an applied external magnetic field. In terms of the \emph{z} component of total spin \emph{S}, two species of TMs have $S_z$=$\pm$1 whereas LM has $S_z$=0. To make $S_z$ a good quantum number, the field direction has to be aligned along the easy-axis and a horizontal-field cryomagnet was employed for that purpose. With an applied field $\mu_0H$, the corresponding Zeeman energy term is g$\mu_{\rm B}$$\mu_0$\emph{H}$S_z$, where g is the Land$\acute{e}$ g-factor and $\mu_{\rm B}$ is the Bohr magneton. Thus the energy shifts of TMs are expected to be $\pm$g$\mu_{\rm B}$$\mu_0$\emph{H} while LM should remain unchanged. Consequently, if the splitting is large enough, LM could be identified by this Zeeman effect.

\begin{figure}
\includegraphics[width=8cm,bbllx=35,bblly=60,bburx=540,bbury=685,angle=0,clip=]{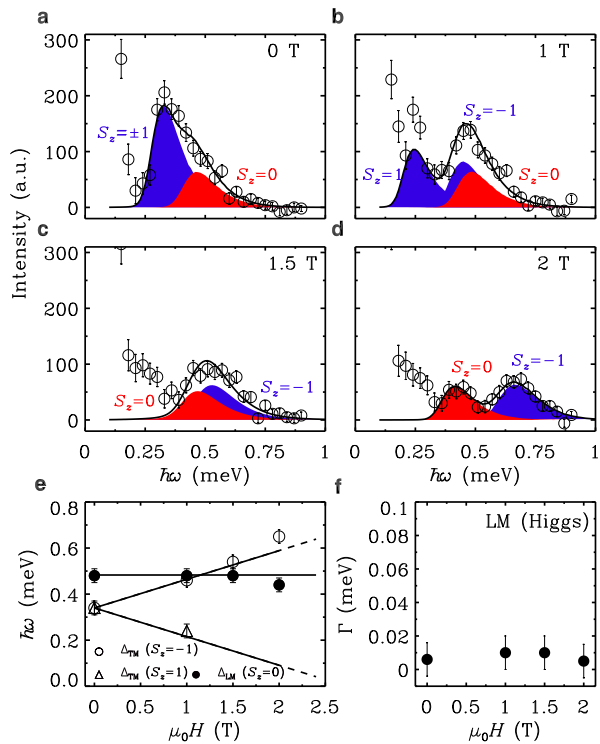}
\caption{\textbf{The Zeeman effect.} The background-subtracted energy scans at the magnetic zone center \textbf{q}=(0.5,-0.5,1.5), \emph{T}=50 mK and (\textbf{a}) $\mu_0H$=0, (\textbf{b}) $\mu_0H$=1 T, (\textbf{c}) $\mu_0H$=1.5 T, and (\textbf{d}) $\mu_0H$=2 T. Solid lines are fits to a double-Lorentzian damped harmonic-oscillator model convolved with the instrumental resolution function and filled areas represent the contribution from the transverse modess (TMs, blue) and the longitudinal mode (LM, red) or the Higgs amplitude mode. \textbf{e}, Field-dependences of TMs ($S_z$=$\pm$1) and LM ($S_z$=0) obtained from the above fits. The dashed lines are the calculated Zeeman energy term as described in the text. The solid lines are the calculations by the bond-operator theory as described in the text. \textbf{f}, Field-dependence of resolution-corrected intrinsic linewidth of the Higgs amplitude mode. Error bars represent one standard deviation determined assuming Poisson statistics.}
\label{fig2}
\end{figure}

Figure~\ref{fig2}a shows the zero-field background-subtracted energy scan at the magnetic zone center \textbf{q}=(0.5,-0.5,1.5) and \emph{T}=50 mK. The spectral lineshape was modeled by superposition of two double-Lorentzian damped harmonic-oscillator (DHO) models convolved with the instrumental resolution function\cite{Hong08:78,Hong10:82}. The best fit yields the gap energies of TMs ($S_z$=$\pm$1) and LM ($S_z$=0) as $\Delta_{\rm TMs}$=0.34(3) meV and $\Delta_{\rm LM}$=0.48(3) meV, respectively. At $\mu_0$\emph{H}=1 T in Fig.~\ref{fig2}b, TMs ($S_z$=$\pm$1) are split into two branches. The observed quasielastic neutron scattering hinders observation of TM ($S_z$=1) at $\mu_0$\emph{H}=1.5 T in Fig.~\ref{fig2}c. At $\mu_0$\emph{H}=2 T in Fig.~\ref{fig2}d, TM ($S_z$=1) is merged into the elastic line while LM becomes clearly visible and well distinguished from TM ($S_z$=-1). Figure~\ref{fig2}e summarizes the measured field-dependences of $\Delta_{\rm TMs}$ ($S_z$=$\pm$1) and $\Delta_{\rm LM}$ ($S_z$=0). $\Delta_{\rm TMs}$ ($S_z$=$\pm$1) as a function of field agree well with the Zeeman spectral splitting $\pm$g$\mu_{\rm B}$$\mu_0$\emph{H} using g=2.15 and LM is indeed field-independent within the experimental uncertainties. The small discrepancy between experimental data and calculation at 2 T is due to occurrence of a spin-flop transition as described in the Supplementary Information. The analysis indicates that the peak profile of the Higgs amplitude mode in each field is limited by the instrumental resolution within experimental uncertainty as shown in Fig.~\ref{fig2}f. In another word, the decay of Higgs amplitude mode ($S_z$=0) into a TM ($S_z$=1) and another TM ($S_z$=-1) is forbidden by the kinematic conditions. However, because of the limited access of the reciprocal space using a horizontal-field cryomagnet, we could not map out the excitation spectra in the Brillouin Zone (BZ).

\begin{figure}
\includegraphics[width=8cm,bbllx=40,bblly=110,bburx=540,bbury=700,angle=0,clip=]{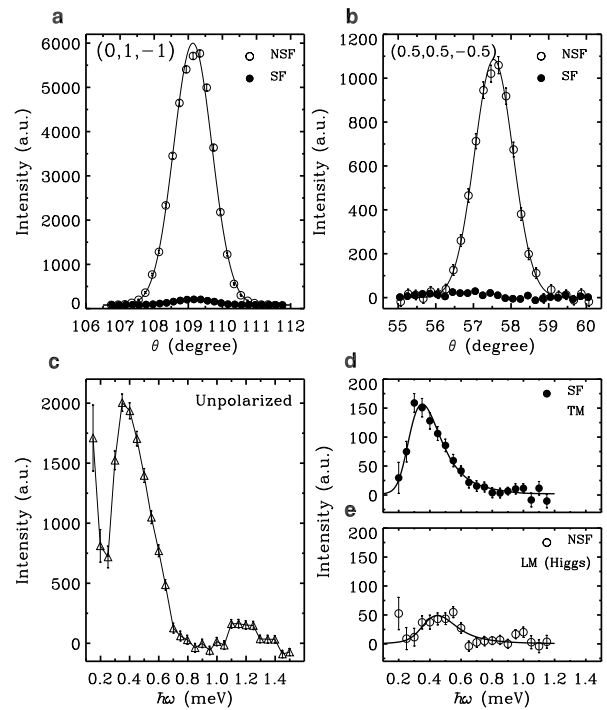}
\caption{\textbf{Feasibility of the polarized neutron study.} Background-subtracted $\theta$-scans at \emph{T}=1.4 K and (\textbf{a}) the nuclear Bragg reflection (0,1,-1) and (\textbf{b}) the magnetic Bragg reflection (0.5,0.5,-0.5) with NSF and SF configurations, respectively. Solid lines are fits to a gaussian function. The background-subtracted transferred energy scans at the magnetic zone center \textbf{q}=(0.5,0.5,-0.5) and \emph{T}=1.4 K with (\textbf{c}) unpolarized neutron, (\textbf{d}) SF, and (\textbf{e}) NSF configurations, respectively. Solid lines are fits to a two-Lorentzian damped harmonic-oscillator model convolved with the instrumental resolution function. Error bars represent one standard deviation determined assuming Poisson statistics.}
\label{fig3}
\end{figure}

Another straightforward way to unambiguously determine the nature of spin polarization of magnetic excitation spectra is by the polarized neutron scattering method. In general, however, it is quite challenging to carry out the polarized INS because of significant loss of neutron intensity due to neutron polarization arrangements compared with unpolarized neutron arrangements. To compensate for that, the polarized neutron data were collected using a high-flux cold neutron spectrometer. The polarization analysis was performed using the recently developed capability of wide-angle $^3$He spin filters\cite{Chen16}. Figure~3a shows the $\theta$-scans of the nuclear Bragg reflection (0,1,-1) at $T$=1.4 K with the non-spin-flip (NSF) and spin-flip (SF) configurations, respectively. The flipping ratio \emph{F} can be calculated as \emph{I}$_{\rm NSF}$/\emph{I}$_{\rm SF}$$\simeq$43(1), which corresponds to an overall polarization efficiency of (\emph{F}-1)/(\emph{F}+1)=0.95. Furthermore, Fig.~3b shows the background-subtracted $\theta$-scans of the magnetic Bragg reflection (0.5,0.5,-0.5) at $T$=1.4 K with NSF and SF configurations, respectively. The scattering intensity in the NSF channel dominates over that in the SF channel, suggesting that the out-of-plane spin component is dominant and thus confirming the determined orientation of the staggered moments.

With the high efficiency of neutron spin filters firmly established and the magnetic spin structure in DLCB well determined, we proceed to investigate the spin dynamics using polarized neutrons. The experiment was designed in such a way that TMs and LM can be separated from each other in SF and NSF configurations, respectively (for further details, see Methods). For the comparison purposes, Fig.~3c shows the background-subtracted energy scan at \textbf{q}=(0.5,0.5,-0.5) and $T$=1.4 K using unpolarized neutrons. Figures~3d,e show the same energy scans with SF and NSF configurations, respectively. And data were fitted to the same DHO model convolved with the instrumental resolution function. The spin gap energies of TMs and LM (Higgs amplitude mode) were obtained as $\Delta_{\rm TMs}$=0.33(3) meV and $\Delta_{\rm LM}$=0.48(3) meV in excellent agreement with the results from the Zeeman effect. The spectral weight ratio between them is about 2.6:1.

\begin{figure}
\includegraphics[width=8cm,bbllx=50,bblly=40,bburx=570,bbury=740,angle=0,clip=]{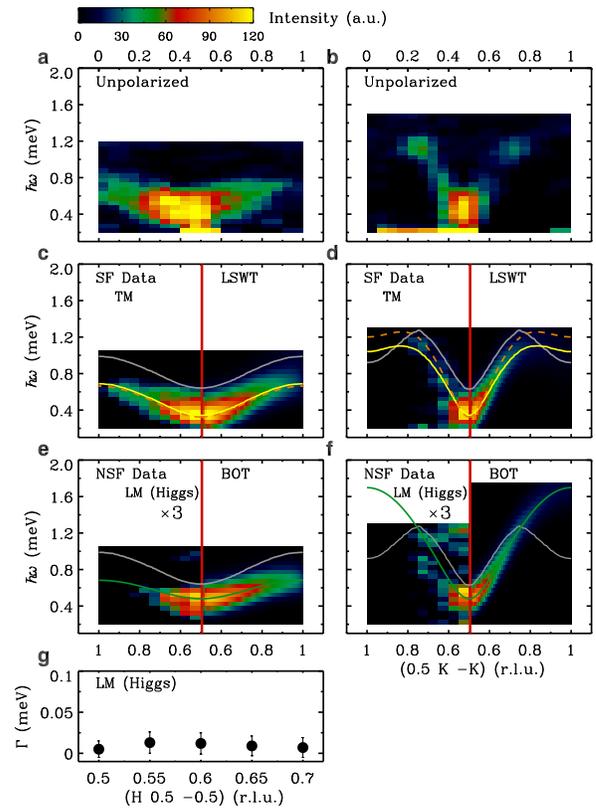}
\caption{\textbf{The polarized neutron study of spin dynamics.} (\textbf{a-b}) False-color maps of the background-subtracted magnetic excitation spectra as a function of energy and wavevector transfer measured using unpolarized neutron; (\textbf{c-d}) Comparison data with SF configuration to the model calculation; (\textbf{e-f}) Comparison data with NSF configuration to the model calculation along the (H,0.5,-0.5) and (0.5,K,-K) directions, respectively. Intensity was enlarged by a factor of 3; (\textbf{g}) the wavevector dependence of the resolution-corrected intrinsic linewidth $\Gamma$ of the Higgs amplitude mode. Note that $\Gamma$ along (0.5 K -K) cannot be reliably obtained due to the weak intensity. Dashed brown lines are the calculation using the high-order series expansions. Solid yellow lines are the calculated transverse modes by LSWT. Solid green lines are the calculated longitudinal (the Higgs Amplitude) modes by BOT. Solid grey lines are the calculated lower-boundary of two-magnon continuum. All experimental data were collected at $T$=1.4 K.}
\label{fig4}
\end{figure}

After confirming the feasibility of such a challenging experiment, we managed to map out the excitation spectra in the BZ. Serving as a reference, Figures~4a,b show the false-color maps of the spin-wave spectra along two high-symmetry directions in the reciprocal lattice space using the unpolarized neutrons. With SF configuration, as expected, the magnetic excitation spectra in Figs.~4c,d are in excellent agreement with the calculations using SPINW\cite{spinw} in the linear spin-wave theory (LSWT) approximation and thus was confirmed as TMs. The Hamiltonian parameters of $J_{\rm leg}$=0.64~meV, $J_{\rm rung}$=0.70~meV, $J_{\rm int}$=0.19~meV, and $\lambda$=0.95 provide the best agreement with experimental data. The modelled dispersion curves in Figs.~4c,d are close to the quantitative calculations using the high-order series expansions\cite{Hong14:89}.

Since the one-magnon excitation of LM is not predicted in LSWT, to analyze the experimental data with NSF configuration, we employed the bond-operator theory (BOT) for the description of the low-energy excitations in the vicinity of QCP on DLCB. The detailed description of the harmonic BOT can be found in Refs.~[\citenum{Sommer-2001,Matsumoto-2002,Matsumoto-2004,Shiina-2003}]. The ordered moment is estimated as 38.5\% of the saturation value and is consistent with the experimental value of 37(5)\%. The exchange interaction parameters were extracted as $J_{\rm leg}$=0.57 meV, $J_{\rm rung}$=1.21 meV, $J_{\rm int}$=0.11 meV, and $\lambda$=0.95 from the best fit. Note that the extracted parameters correspond to the renormalized ones within the scope of harmonic BOT. Since DLCB is a weakly interacting ladder system, $J_{\rm rung}$ and $J_{\rm leg}$ are strongly renormalized (the former and latter are enhanced and reduced, respectively)\cite{Gopalan-1994,Normand-2011}. Therefore, $J_{\rm rung}$ is markedly larger than $J_{\rm leg}$. The solid green lines in Figs.~4e,f show the calculated LM (Higgs amplitude mode) with a gap energy at 0.48 meV. We notice that the BOT calculation in Fig.~4f monotonically increases and deviates from the experimental data at the zone boundary, which may originate from the fact that the BOT is a mean-field treatment and application of this technique to the low-dimensional system could be limited. For the low-energy excitations, nevertheless, it works well in both QD and LRO phases. For instance, as shown in Fig.~2e the agreement of the field dependence of Zeeman energy term between the experimental data and the calculations by BOT is excellent.

Figures 4e,f show the calculated excitation spectra by BOT, which reproduce the experimental data qualitatively. Thus, our conclusion that the nature of spin excitation observed in NSF configuration is due to spin fluctuation along the staggered moment direction is quite convincing. Note that the Higgs amplitude mode in DLCB is distinctly different from the longitudinal excitations in the \emph{S}=1/2 2D Heisenberg square-lattice (HSL) antiferromagnet Cu(DCOO)$_2$$\cdot$4D$_2$O (CFTD)\cite{Christensen-2007,Dalla-2015} because: (i) The \emph{S}=1/2 2D HSL antiferromagnet is far from QCP. (ii) The observed longitudinal spectra in CFTD originate from the two-magnon continuum. Consequently, the spectral lineshapes are broadened. (iii) Recent theoretical work\cite{Powalski-2015,Powalski-2017} suggests that there is a prominent resonance, which was proposed as a Higgs resonance with finite lifetime, inside the continuum due to the attractive magnon-magnon interaction. Moreover, the grey lines in Figs. 4e,f are the calculated lower boundary of the two-magnon continuum in DLCB, which lies well above the Higgs amplitude mode. Hence, the spontaneous magnon decay of the Higgs amplitude mode into a pair of TMs is forbidden due to violation of the laws of energy and momentum conservation as evidenced by the wavevector dependence of the intrinsic linewidth $\Gamma$, which is limited by the instrumental resolution as shown in Figs. 4g.

In summary, the unique ability of neutron scattering to probe the spin polarization of dynamic
spin pair-correlation functions allows to distinguish the Higgs amplitude mode from the dominant transverse Nambu-Goldstone mode in a two-dimensional \emph{S}=1/2 antiferromagnet DLCB. The transverse modes have a finite excitation gap energy due to a weak Ising anisotropy. The opening of the gap kinematically prevents the decay process from the Higgs amplitude mode to a pair of transverse modes. This leads to the long lifetime for the Higgs amplitude mode and makes it observable near the quantum critical point in two-dimensions.

\emph{Note added}.---Recently we became aware of an INS work\cite{Jain17} that reports the Higgs amplitude mode in a 2D antiferromagnet Ca$_2$RuO$_4$.

\section*{\large Methods}
\noindent
\textbf{Single crystal growth.} Deuterated single crystals were grown using a solution method\cite{Awwadi08:47}. An aqueous solution containing a 1:1:1 ratio of deuterated (DMA)Br, (35DMP)Br, where DMA$^+$ is the dimethylammonium cation and 35DMP$^+$ is the 3,5-dimethylpridinium cation, and copper(II) bromide was allowed to evaporate over several days in a closed desiccator. A few drops of DBr were added to the solution to avoid hydrolysis of the Cu(II) ion.\\

\noindent
\textbf{Neutron scattering measurements.} Unpolarized neutron-scattering measurements using a horizontal-field superconducting magnet with a dilution fridge insert were carried out on a cold neutron triple-axis spectrometer (FLEXX)\cite{Le13:729} at Helmholtz-Zentrum Berlin. The sample consists of two co-aligned deuterated single crystals with a total mass of 2.5 g and a 1.0$^\circ$ mosaic spread and was oriented in the (H -H L) reciprocal-lattice plane. Unpolarized inelastic neutron-scattering measurements using a standard Helium-flow cryostat were carried out on a cold neutron triple-axis spectrometer (CTAX) at High Flux Isotope Reactor, Oak Ridge National Laboratory. Polarized neutron-scattering measurements using a standard Helium-flow cryostat were performed on a high intensity multi-axis crystal spectrometer (MACS)\cite{Rodriguez08:19} at the NIST Center for Neutron Research. The peak flux at sample position is about 5$\times$10$^8$ neutrons/cm$^2$/s. The sample assembly with three co-aligned deuterated single crystals (a total mass of 3.5 g and a 1.0$^\circ$ mosaic spread) was oriented in the (H K -K) reciprocal-lattice plane. In all experiments, the final neutron energy was fixed at 3.0 meV and the energy resolution of FLEXX and MACS at the elastic line are 0.10 meV and 0.15 meV, respectively. The background was determined at \emph{T}=15 K under the same instrument configuration and has been subtracted.\\

\noindent
\textbf{Polarized neutron measurements.} In the experimental setup, both incident and outgoing neutron beams were polarized by nuclear spin polarized $^3$He gas cells. NMR-based inversion of the $^3$He polarization in the polarizer cell allows polarization of the incident beam parallel or antiparallel to the vertical axis at will. The overall transmission at the beginning of a polarized neutron setup (each run lasts about two days) is about 11$\%$ and reduced to about 5$\%$ before the $^3$He gas cells change out. The initial flipping ratio \emph{F} is about 43(1), indicating that the product of the polarizing efficiency of the NSF cells \emph{PA}=(\emph{F}-1)/(\emph{F}+1) is 95$\%$. Typically, after the two-day operation, \emph{F} is reduced to 20(1) and \emph{PA} becomes 91$\%$. Since \emph{F} was always above 20, the polarization leakage effect is as small as 1/\emph{F} ($\leq$5$\%$). To account for decay of the $^3$He polarization and neutron transmission with time, the polarized neutron data were corrected by $^3$He efficiency correction software as described in Supplementary Information.

The principles for polarized neutron scattering can be summarized as follows: a. Phonons and structural scattering are seen in NSF channel; b. components of spin fluctuations parallel to the direction of neutron polarization are seen in the NSF channel;  c. components of spin fluctuations perpendicular to the direction of the neutron polarization are seen in SF channel\cite{Moon69:181}.

The sample was aligned in such a way that the [0,1,1] direction in the real space is vertical. Thus, the angle $\alpha$ between the vertical polarization and staggered moment direction is 17.6$^\circ$. In this geometry with the NSF configuration, the large fraction ($\cos^2$$\alpha$$\simeq$91$\%$) is due to the spin fluctuations along the direction of the staggered moment (LM or the Higgs amplitude mode) while the remaining 9$\%$ corresponds to the spin fluctuations perpendicular to the staggered moment (TMs) and is negligible. In contrast, in the SF configuration, TMs have accounted for 91$\%$ of the contribution and LM or the Higgs amplitude mode is negligible. Therefore, by employing polarized INS, we are able to separate the LM (the Higgs amplitude mode) from the TMs in the magnetic excitation spectra. Polarized neutron measurements cover half of Brillouin Zone due to the fact that the polarizing efficiency becomes either significantly reduced or not available for small neutron scattering angles.\\

\noindent
\textbf{Data analysis.} The spectral lineshape in Figs.~2(a)-(d) and Figs.~3(d)-(e) was fitted to the following double-Lorentzian damped harmonic-oscillator model
\begin{eqnarray}
S(\hbar\omega)=
\frac{A}{1-\exp(-\hbar\omega/k_{\rm B}T)} \\ \nonumber \left[ \frac{\Gamma}{(\hbar\omega-\Delta)^2+\Gamma^2} \right.
  - \left. \frac{\Gamma}{(\hbar\omega+\Delta)^2+\Gamma^2}
\right],\nonumber
\label{sqw}
\end{eqnarray}
where $\Delta$ is the peak position and $\Gamma$ is the resolution-corrected intrinsic excitation linewidth i.e. half-width at half-maximum (HWFM), and convolved with the instrumental resolution function. The experimental resolution was calculated using the Reslib software\cite {reslib}.

For the false-color maps in Figs.~4(a)-(f), data were obtained by combining a series of constant-\textbf{q} scans along the either (H 0.5 -0.5) or (0.5 K -K) direction with step size of 0.05 r.l.u. and simulations were convolved with the instrumental resolution function where the neutron polarization factor and the magnetic form factor for Cu$^{2+}$ were included.

Detailed description about the determination of the lower boundary of two-magnon continuum can be found in Ref.~[\citenum{Hong17:08}].\\

\noindent
\textbf{Data availability}. The data that support the plots within this paper and other findings of this study are available from the corresponding author upon request.

\section*{\large Acknowledgments}
T.H. thanks C. D. Batista for the insightful discussion and R. Erwin for the development of $^3$He efficiency correction software. T.H. also thanks D. L. Quintero Castro, Z. L. Lu, and Z. H$\rm \ddot{u}$sges for the assistance during the experiment. One of the authors (M.M.) is supported by JSPS KAKENHI Grant Number 26400332. A portion of this research used resources at the High Flux Isotope Reactor, a DOE Office of Science User Facility operated by the Oak Ridge National Laboratory. The work at NIST utilized facilities supported by the NSF under Agreement No. DMR-1508249.\\

\section*{\large Author contributions}
T.H. conceived the project. F.F.A. and M.M.T. prepared the samples. The polarization apparatus and corrections were provided by W.C, T.G. and S.W. T.H., Y.Q., H.A., R.T., and B.K. performed the neutron-scattering measurements. T.H., M.M., D.A.T., S.E.D., K.C., and K.P.S. analysed the data. All authors contributed to the writing of the manuscript.\\

\section*{\large Additional information}
\noindent
\textbf{Competing financial interests:} The authors declare no competing financial interests. \\ \\
\textbf{Materials \& Correspondence.} Correspondence and requests for materials should be addressed to T.H. (email: hongt@ornl.gov).

\noindent
\section{\large \textbf{Supplementary Information}}

\beginsupplement

\section{MEASUREMENT OF THE INTER-LAYER DISPERSION}
\begin{figure}
\includegraphics[width=8cm,bbllx=75,bblly=45,bburx=465,bbury=685,angle=0,clip=]{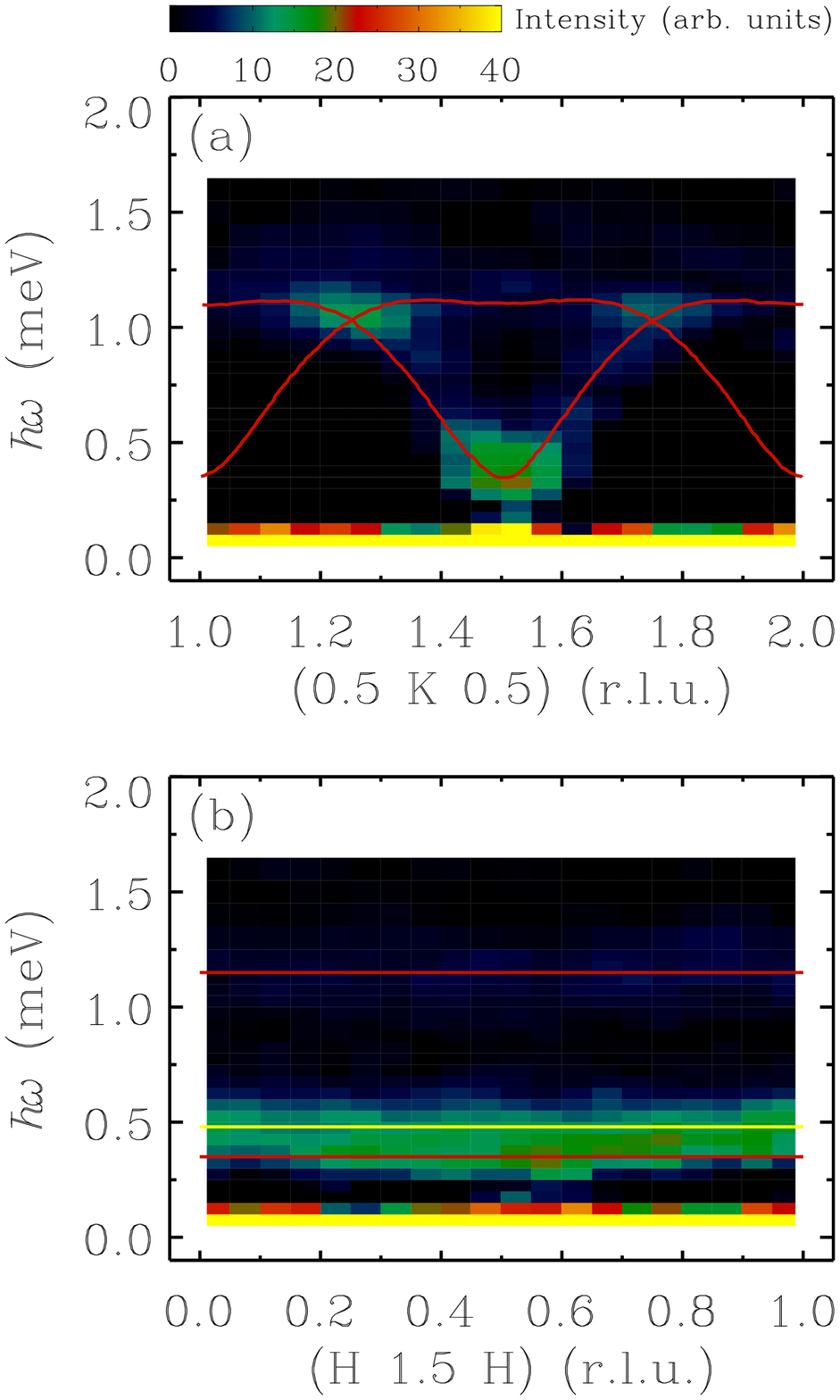}
\caption{Background subtracted magnetic excitation spectra as a function of energy and wavevector transfer measured at \emph{T}=1.5 K along (\textbf{a}) the leg direction and (\textbf{b}) the inter-layer direction. The red and yellow lines are the calculations by the linear spin-wave theory and the bond-operator theory, respectively, as described in the main text. The slice along (0.5 K 0.5) is integrated in the range 0.4$<$H$<$0.6. The slice along (H 1.5 H) is integrated in the range 1.4$<$K$<$1.6.}
\label{s1}
\end{figure}

Additional unpolarized inelastic neutron scattering measurements were performed at the neutron spectrometer MACS to investigate the possible interaction between the two-dimensional layers in DLCB. For that purpose, a single crystal ($\sim$2 g) with 1.0$^\circ$ mosaic spread was aligned in the (H K H) reciprocal-lattice plane. The final neutron energy was fixed at 3.0 meV. The background was determined at \emph{T}=15 K under the same instrument configuration and has been subtracted. Figure~\ref{s1}a shows the measured excitation spectrum at \emph{T}=1.5 K along the leg direction and the observed dispersions are fully consistent with the calculations by the linear spin-wave theory as described in the main text. Along the inter-layer direction as shown in Fig.~\ref{s1}b, the dispersion is absent within the instrumental resolution indicating that DLCB is an excellent two-dimensional system.

\section{ANALYSIS OF THE SPIN GAP ENERGY BY TUNING THE INTER-RUNG INTERACTIONS}
\begin{figure}
\includegraphics[width=8cm,bbllx=35,bblly=10,bburx=600,bbury=425,angle=0,clip=]{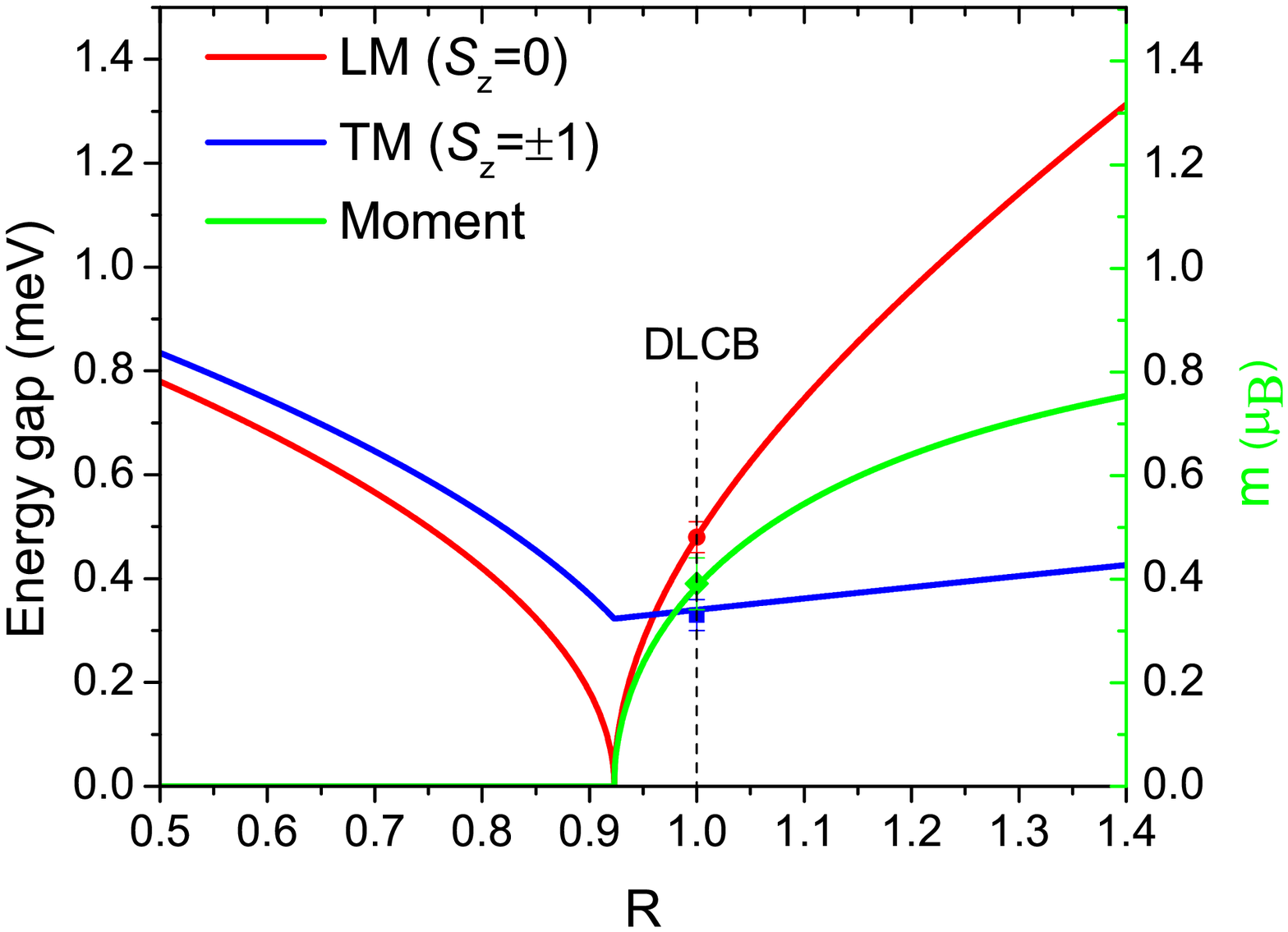}
\caption{Calculations of the spin gap energy of the triplet modes with an Ising anisotropy (left) and the staggered moment size (right) by the bond-operator theory as a function of an enhancement factor \emph{R} for the inter-rung interactions. A quantum critical point is at $R_c$=0.923 while DLCB is at \emph{R}=1. Circle, diamond, and square points are the observed values from the experimental measurements.}
\label{s2}
\end{figure}

We proposed a two-dimensional model for magnetic interactions as shown in Fig.~1b in the main text. Due to the Ising anisotropy, the triplet spin gap energy splits into a singlet ($S_z$=0) and a doublet ($S_z$=$\pm$1). We assume that $J_{\rm rung}$ is fixed and the inter-rung interactions vary as $J_{\rm leg}^*$=\emph{R}$J_{\rm leg}$ and $J_{\rm int}^*$=\emph{R}$J_{\rm int}$, where $J_{\rm rung}$, $J_{\rm leg}$, and $J_{\rm int}$ are values obtained in DLCB. Figure~\ref{s2} summaries the calculations by the bond-operator theory (BOT) of the evolution of the spin gap energy as a function of the enhancement factor \emph{R}. When \emph{R} initially increases, the spin gap energies of both the singlet and the doublet decrease. Spin gap of the singlet closes at the quantum critical point (QCP) while the doublet remains gapped. When \emph{R} is further increased, the softened singlet mode acquires a spin gap again and becomes the Higgs amplitude mode. The analysis indicates that QCP locates at $R_c$=0.923, which is close the case in DLCB (\emph{R}=1) and thus confirms our conclusion in the main text. For $R<R_c$, the quantum disordered phase is stablized while the long-range ordered phase is stabilized for $R>R_c$. Calculation by BOT of emergence of the staggered moment size as a function of \emph{R} is also shown in Fig.~\ref{s2}.

\section{SPIN-FLOP TRANSITION}
In DLCB, we have shown that the staggered moments point along an easy axis ($\equiv$$\hat{z}$). An application of a magnetic field along the direction of the ordered moment would lead to the spin reorientation i.e. spin-flop transition. In the spin-flop phase, \emph{z} component of total spin \emph{S} ($S_z$) is no longer a good quantum number. To find out the critical field where the spin reorientation occurs, we measured several magnetic reflections, which can be accessed by the limited reciprocal lattice space of the horizontal magnet, using the spectrometer. The sample was oriented in the (H -H L) scattering plane and the field direction was applied parallel to the easy-axis of DLCB. Both the incident and the final neutron energies were fixed at 4.0 meV.

\begin{figure}
\includegraphics[width=8cm,bbllx=65,bblly=60,bburx=465,bbury=715,angle=0,clip=]{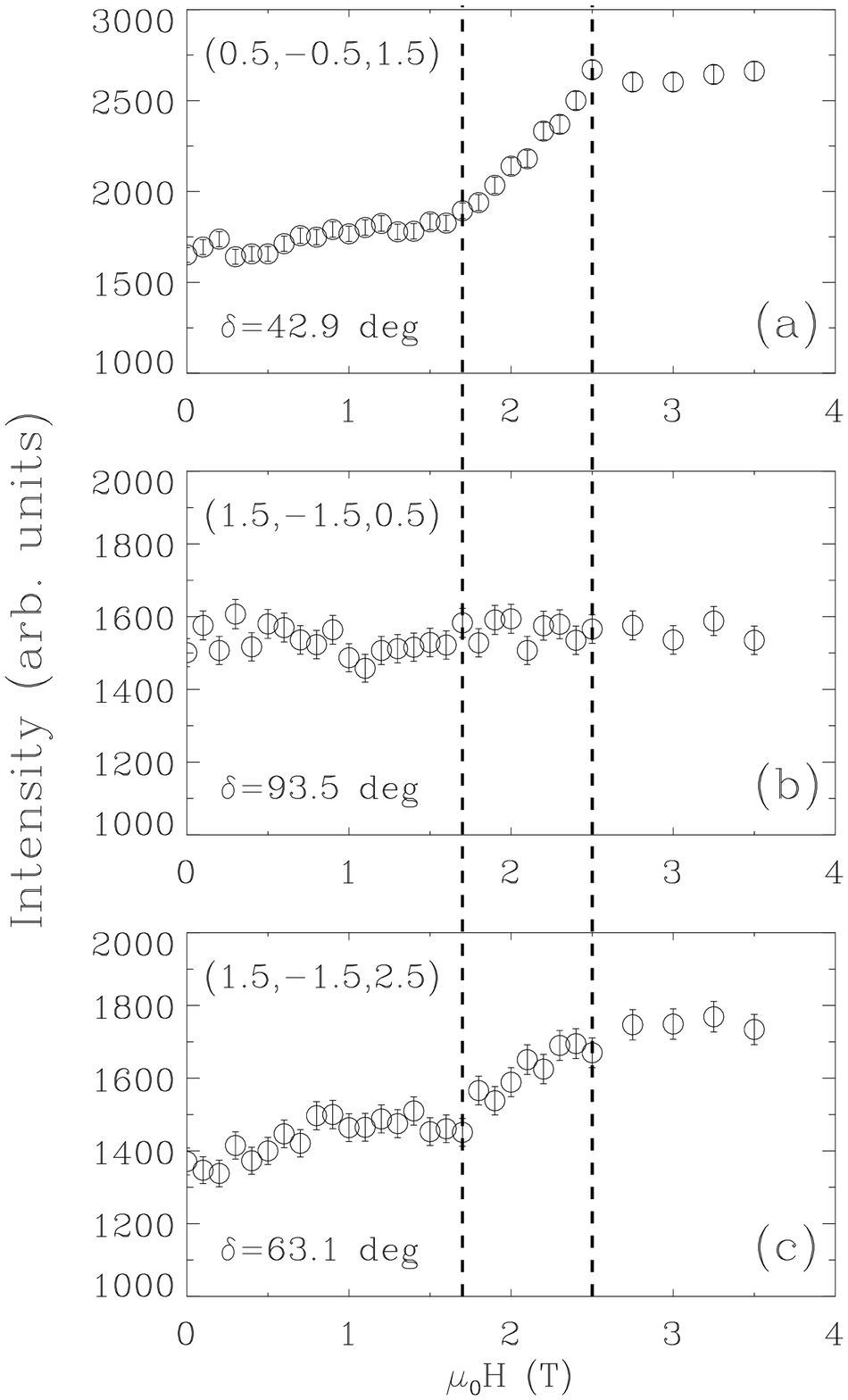}
\caption{Magnetic field dependencies of the integrated-peak intensity of several magnetic Bragg reflections (\textbf{a}) \textbf{q}=(0.5,-0.5,1.5), (\textbf{b}) \textbf{q}=(1.5,-1.5,0.5), and (\textbf{c}) \textbf{q}=(1.5,-1.5,2.5) in DLCB measured at $T$=50 mK. $\delta$ is the angle between direction of the ordered moment and the tranferred wavevector. The dashed lines define the spin-canted region. Error bars represent one standard deviation determined assuming Poisson statistics.}
\label{s3}
\end{figure}

Figure~\ref{s3} shows the measured field dependencies up to 3.5 T of several magnetic Bragg peaks. The integrated-peak intensities are almost field-independent at low fields, then start to increase above 1.7 T for the (0.5,-0.5,-1.5) and (1.5,-1.5,2.5) magnetic reflections, and become saturated above 2.5 T after the spin is flopped. From the fact that neutron scattering probes the components of spin fluctuation perpendicular to the transferred wavevector, the orientation of the ordered moment in the spin-flop phase is 90 degree out of the horizontal plane with the axis of rotation along the transferred wavevector (1.5,-1.5,0.5).

Overall, the above results confirm that in DLCB when the field direction is aligned parallel to the staggered moment direction, $S_z$ remains as a good quantum number at least up to 1.7 T.

\section{POLARIZED NEUTRON BEAM CALIBRATION AND POLARIZATION EFFICIENCY CORRECTION}

\subsection{Polarized $^{3}$$\rm\bf{He}$ as a neutron spin filter}
   \label{principle}
A \tHe\  neutron spin filter is a transmission-based neutron-polarizing device\cite{Jones00}.  The neutron transmission through a polarized \tHe \ cell for each spin state is given by
\begin{equation}
	T^{\pm}=T_{\mathrm E}{\mathrm {exp}}\left[- \mathcal{O}\left(1\mp P_{\mathrm {He}}\right)\right]\label{T+-}
\end{equation}
where $T^{\pm}$ is the transmission with neutron spin parallel ($+$) or antiparallel ($-$) to the \tHe\ spin. $T_{\mathrm E}$ is the transmission of the \tHe\ glass cell windows, and $P_{\mathrm {He}}$ is the \tHe\ polarization. $\mathcal{O}$ is the opacity (gas thickness) of the cell, which is linearly proportional to the neutron wavelength, \tHe\ gas density and the neutron path length through the cell. The transmission for an unpolarized neutron beam passing through a polarized \tHe\ cell is then given by
\begin{equation}
	T_{\mathrm n}=T_{\mathrm E}{\mathrm {exp}}\left(- \mathcal{O}\right) {\mathrm {cosh}} \left(\mathcal{O} P_{\mathrm {He}}\right)=T_{0}{\mathrm {cosh}} \left(\mathcal{O} P_{\mathrm {He}}\right) \label{Tn}
\end{equation}
where $T_{0}$ is the transmission for an unpolarized neutron beam passing through an unpolarized \tHe\ cell ($P_{\mathrm {He}}$=0) and is given by
\begin{equation}
   T_{0}=T_{\mathrm E} {\mathrm {exp}}\left(-\mathcal{O}\right) \label{T0}
\end{equation}
The resulting neutron polarization after passing through a polarized \tHe\ cell is given by
\begin{equation}
	P_{\mathrm n}=\frac{T^{+}-T^{-}}{T^{+}+T^{-}}={\mathrm {tanh}}\left(\mathcal{O} P_{\mathrm {He}}\right) = \sqrt{1-\frac{T_{0}^2}{T_{\mathrm n}^2}}\label{T+-}
\end{equation}

The \tHe\ polarization for each spin filter (polarizer and analyzer) can be individually determined from measurements of $T_{0}$, $T_{\mathrm n}$ and $T_{\mathrm E}$ using Eqs.~(\ref{Tn}) and (\ref{T0}).

\subsection{polarization efficiency correction} \label{PECor}

The polarized beam apparatus on MACS contains a \tHe\ polarizer before the sample and two wide-angle \tHe\ spin analyzers after the sample\cite{Fu11,Ye13}. The neutron spin state of the incident beam can be flipped by inverting the \tHe\ polarization in the polarizer using the nuclear magnetic resonance (NMR) method. For this experiment, only one \tHe\ analyzer was used. The performance of the \tHe\ polarizers and analyzers on MACS have recently been improved significantly so that a sample with a small magnetic moment could be measured successfully\cite{Chen16}. Since the neutron polarizing efficiencies of these  \tHe\ polarizers and analyzers are less than 100~\%, a correction to the raw data for the small leakage of the minority spin state  is necessary before further data analysis is done\cite{Majkrzak91, Wildes06}.

Here we briefly describe the procedure for the polarization efficiency correction. A complete description of polarized beam calibration and polarization efficiency correction will be presented in an upcoming publication. The neutron spin states of the incident and scattered beams can be either \textquotedblleft up\textquotedblright or \textquotedblleft down\textquotedblright, hence four possible spin dependent cross sections, $I^{\mathrm {uu}}$, $I^{\mathrm {ud}}$, $I^{\mathrm {du}}$, and $I^{\mathrm {dd}}$, can be measured, where the first (second) letter, either \textquotedblleft u\textquotedblright or \textquotedblleft d\textquotedblright, refers to the spin state of the incident and scattered beam, respectively. $I^{\mathrm {uu}}$ and $I^{\mathrm {dd}}$ are referred to as non-spin-flip scattering, $I^{\mathrm {ud}}$ and $I^{\mathrm {du}}$ are referred to as spin-flip scattering. The neutron spin flip efficiency is close to 100~\% and is neglected here since the \tHe\ polarization loss is less than 1x10$^{-3}$ per flip\cite{Ye13}. The probability of the neutron being depolarized by the sample is denoted as $\epsilon$. For this sample, it is valid to assume that ${\sigma^{\mathrm {uu}}}  = {\sigma^{\mathrm {dd}}}$ and ${\sigma^{\mathrm {du}}}   = {\sigma^{\mathrm {ud}}}$, hence the four possible spin-dependent cross sections are reduced to two. In general we measure $I^{\mathrm {uu}}$ and $I^{\mathrm {du}}$.

Now we can write two measured spin-dependent intensities in a linear combination of the non-spin flip ($\sigma^{\mathrm {uu}}$) and spin flip ($\sigma^{\mathrm {du}}$) scattering from the sample,

\begin{eqnarray}
\label{iuu_mod}
I^{\mathrm {uu}}=\left[ \left(T_{\mathrm {P}}^{+}T_{\mathrm {A}}^{+} + T_{\mathrm {P}}^{-}T_{\mathrm {A}}^{-}\right)\left(1-\epsilon\right)  +  \left(T_{\mathrm {P}}^{+}T_{\mathrm {A}}^{-} + T_{\mathrm {P}}^{-}T_{\mathrm {A}}^{+}\right)\epsilon \right] \sigma^{\mathrm {uu}} + \nonumber \\
 \left[ \left(T_{\mathrm {P}}^{+}T_{\mathrm {A}}^{+} + T_{\mathrm {P}}^{-}T_{\mathrm {A}}^{-}\right)\epsilon  +  \left(T_{\mathrm {P}}^{+}T_{\mathrm {A}}^{-} + T_{\mathrm {P}}^{-}T_{\mathrm {A}}^{+}\right)\left(1-\epsilon\right) \right] \sigma^{\mathrm {ud}} + B
 \end{eqnarray}
\begin{eqnarray}
\label{idu_mod}
I^{\mathrm {du}}=\left[ \left(T_{\mathrm {P}}^{+}T_{\mathrm {A}}^{+} + T_{\mathrm {P}}^{-}T_{\mathrm {A}}^{-}\right)\epsilon  +  \left(T_{\mathrm {P}}^{+}T_{\mathrm {A}}^{-} + T_{\mathrm {P}}^{-}T_{\mathrm {A}}^{+}\right)\left(1-\epsilon\right) \right] \sigma^{\mathrm {uu}}  + \nonumber \\
 \left[ \left(T_{\mathrm {P}}^{+}T_{\mathrm {A}}^{+} + T_{\mathrm {P}}^{-}T_{\mathrm {A}}^{-}\right)\left(1-\epsilon\right)  +  \left(T_{\mathrm {P}}^{+}T_{\mathrm {A}}^{-} + T_{\mathrm {P}}^{-}T_{\mathrm {A}}^{+}\right)\epsilon \right] \sigma^{\mathrm {du}} + B
 \end{eqnarray}

where $T^{+}$ and $T^{-}$ are defined earlier and the subscript \textquotedblleft P\textquotedblright or \textquotedblleft A\textquotedblright refers to as either the polarizer or the analyzer. $B$ is the instrument background, and was determined to be negligible. $\epsilon$ can be determined from the difference in the flipping ratios between a pyrolytic graphite sample and the real sample. For this experiment, $\epsilon$ was determined to be negligible.

The \tHe \ polarization decayed exponentially with its own characteristic relaxation time, $T_{1}$, for each polarizer and analyzer.  $T_{1}$s were obtained from the initial and final \tHe \ polarization measurements for each polarizer and analyzer cell, which were compared to FID (free induction decay) NMR measurements during the experiment. $T_{1}$s determined from the two methods agreed well with each other.  All coefficients in Eqs.~(\ref{iuu_mod}) and (\ref{idu_mod}) were determined from the initial \tHe \ polarizations and $T_1$s of both the polarizer and analyzer at any given time during the experiment. $\sigma^{\mathrm {uu}}$ and $\sigma^{\mathrm {du}}$ from the sample were then solved for any given data point. This procedure was repeated for the raw data set at 1.4 K and 15 K. Following these polarization efficiency corrections, the resultant non-spin flip and spin flip scattering from the sample were obtained by subtracting the 15 K data from the 1.4 K data.

The polarizer and analyzer cells were polarized in an off-line facility for spin-exchange optical pumping\cite{Chen14} that has recently been upgraded and expanded for polarized beam applications on MACS.  After transport to the neutron beam line, the initial \tHe \ polarizations were between 82\% and 85\% and between 70\% and 79\% for the \tHe \ polarizer cells and analyzer cells, respectively.  $T_{1}$s were between 180~h and 260 h and between 160~h and 370~h for the \tHe \ polarizer cells and analyzer cells, respectively. Hence the flipping ratio decayed from about 44 to about 20 after two days. As a result, the correction to the leakage of the minority spin state was only between 2\% and 5\% during the two-day experiment. The overall transmission of the polarized beam configuration as compared to the unpolarized beam configuration was about 11\% with freshly polarized \tHe \ polarizer and analyzer cells, and decayed to between 4\% and 5\% after two days.

\end{document}